\documentclass[aps,showpacs,prb,superscriptaddress,floatfix,twocolumn,citeautoscript]{revtex4-1}

\usepackage{graphicx}
\usepackage{amsmath}
\usepackage{amssymb}
\usepackage{float}
\usepackage{bm}
\usepackage{dcolumn}
\usepackage{subfigure} 
\usepackage{hyperref}
\usepackage[usenames]{color}
\usepackage{multirow}
\usepackage{subfigure}
\usepackage{lipsum}

\hypersetup{pdfborder=0 0 0,colorlinks=true,citecolor=blue,linkcolor=blue}
\input epsf
\DeclareGraphicsExtensions{.pdf,.jpeg,.png, .eps}

\begin{document}
	
\title{Plasmon-phonon coupling in a valley-spin-polarized two-dimensional electron system: a theoretical study on monolayer silicene}

\author{M. Mirzaei}
\affiliation{Department of Physics, Shahid Beheshti University, G. C., Evin, Tehran 1983969411, Iran}
\author{T. Vazifehshenas}
\email{t-vazifeh@sbu.ac.ir}
\affiliation{Department of Physics, Shahid Beheshti University, G. C., Evin, Tehran 1983969411, Iran}
\author{T. Salavati-fard}
\affiliation{Department of Physics and Astronomy, University of Delaware, Newark, DE 19716, USA}
\affiliation{Department of Chemical and Biomolecular Engineering, University of Houston, Houston, TX 77204, USA}
\author{M. Farmanbar}
\affiliation{Faculty of Science and Technology and MESA+ Institute for Nanotechnology, University of Twente, P.O. Box 217, 7500 AE Enschede, The Netherlands}
\author{B. Tanatar}
\affiliation{Department of Physics, Bilkent University, Bilkent, 06800 Ankara, Turkey}

\begin{abstract}
	We study the hybrid excitations due to the coupling between surface optical phonons of a polar insulator substrate and plasmons in the valley-spin-polarized metal phase of silicene under an exchange field. We perform the calculations within the generalized random-phase approximation where the plasmon-phonon coupling is taken into account by the long-range Fr$\ddot{\mathrm{o}}$hlich interaction. Our investigation on two hybridized plasmon branches in different spin and valley subbands shows distinct behavior compared to the uncoupled case. Interestingly, in one valley, it is found that while the high energy hybrid branch is totally damped in the spin-up state, it can be well-defined in the the spin-down state. Moreover, we show that the electron-phonon coupling is stronger in both spin-down subbands, regardless of valley index, due to their higher electron densities. In addition, we study the effects of electron-phonon coupling on the quasiparticle scattering rate of four distinct spin-valley locked subbands. The results of our calculations predict a general enhancement in the scattering rate for all subbands, and a jump in the case of spin-down states. This sharp increase associated to the damping of hybrid plasmon modes is almost absent in the uncoupled case. The results suggest an effective way for manipulating collective modes of valley-spin-polarized silicene which may become useful in future valleytronic and spintronic applications. 
	
\end{abstract}



\maketitle

\section{Introduction}
Recently, the emergence of valley electronics has stimulated a lot of research interest, both theoretically and experimentally. Silicene-based devices are predicted to be potential candidates for valleytronics applications. The term valleytronics refers to the manipulation and utilization of the electron valley index to store and carry information. The valley
which is the local energy extremum in the band structure of honeycomb lattices with two inequivalent Dirac points, has a definite chirality due to the pseudospin-orbit coupling. \cite{valley,valley2} Therefore, the valley can be considered as an extra degree of freedom and consequently, valley-dependent physics is relevant for these structures. 

Silicene is a monolayer honeycomb lattice of silicon with a slightly buckled structure\cite{silicene,synthes1,synthes2,synthes3,borowik2016monte,zhao2016rise}. The inherent buckling results in the generation of an on-site electric potential difference between two sublattices of silicene, upon perpendicularly applying an external electric field. This on-site potential difference, however, is responsible for breaking the spin degeneracy of the energy subbands.\cite{EzEzawa,bandgap1,bandgap2} Carriers in silicene, and some other honeycomb materials such as transition metal dichalcogenides and germanene possess the  valley degree of freedom.  
On the other hand, unlike graphene, silicene has a finite band gap due to the intrinsic spin-orbit coupling (SOC)\cite{bandgap1, spin1,soc} that can be electrically controlled. Interestingly, it is possible to separate the inequivalent Dirac points in silicene, through applying an external fields, and investigate tunable valley-dependent transport properties\cite{peters,tabert} in such systems. 
The applied electric field can cause opposite effects in the two valleys, for example, by tuning the electric field $E_{z}$, the spin-up band gap may be increased and the gap between the spin-down subbands at the same valley may be reduced while the changes are totally reversed in the other valley. 

In a special but important case, when the applied electrostatic potential energy is equal to the SOC energy, the system enters the valley-spin-polarized metal (VSPM) phase in which one of the spin band gaps vanishes and the other remains open in one valley, whereas the situation is completely the other way around in the other valley. This valley-spin locked state can also be obtained by photo-irradiation. 
Moreover, the exchange field, $M$, that can be induced by adatoms or the ferromagnetic substrates, is capable of breaking the time reversal symmetry of the system and producing the Zeeman effect.\cite{peters} As a result, the subband density of states and chemical potential will depend on the exchange field.
According to the above-mentioned properties, several interesting phenomena such as the quantum spin Hall and quantum valley Hall conductances can be observed in silicene.\cite{VHE,VHE1,VHE2,soc} 

There are several electron scattering mechanisms in such a system, among which the electron-electron interaction that gives rise to collective plasmon oscillations and single-particle excitations, is of great importance. Using the random-phase approximation (RPA) for electron-electron interaction, the plasmon dispersion of silicene has been calculated and the variation of the plasmonic resonance modes with the perpendicularly applied electric field has been examined.\cite {tabert} It is found that the behavior of undamped plasmons is strongly dependent on the strength of the electric field and the location of chemical potential with respect to the band gaps. In another interesting theoretical work, the effects of both applied exchange and electric fields on the electron-electron interaction have been investigated in silicene and analytical and numerical results for the dispersion, lifetime and oscillator strength of plasmons have been obtained. As a result of combined effects of both fields, the single-particle excitation (SPE) spectrum acquires a valley and spin texture, thus the valley- and spin-polarized plasmons are predicted. \cite{peters}  

In addition to the electron-electron coupling, the scattering of electrons by the underlying substrate may be significant in the supported silicene structure. As a matter of fact, producing the freestanding silicene is still a challenging open issue, so formation of silicene on several metallic and non metallic substrates has been investigated by many researchers. In particular, silicene has been synthesized on Ag(111) frequently and on ZrB$_2$(0001)\cite{synthes3}, ZrC(111)\cite{zrc28}, Ir(111)\cite{ir29} and MoS$_2$\cite{mos230}. At the same time, there have been some computational studies on designing the non-metallic substrates for silicene, as well\cite{substrate, houssa2014interaction}.  
Interestingly, fabrication of the first silicene field effect transistor by placing an encapsulated silicene (between Al$_2$O$_3$ and Ag) on SiO$_2$ substrate\cite{tao}, provides hope for transferring silicene on other conventional insulator substrates. 

A polar substrate can affect the transport properties such as mobility, an important parameter for micro- and nano-scale field effect transistors, through coupling between the surface optical phonons (SO phonons) of substrate and the electrons of silicene.\cite{kokott2014nonmetallic, chen2016designing} In another words, the longitudinal surface optical phonons of the polar insulator substrate generates a long-range electric field that influences the transport of charged carriers, notably.\cite{Mahan} The remarkable interaction between electrons and SO phonons, is beyond the single-particle properties and comes mainly from the plasmon-SO phonon coupling. This long-wavelength effect leads to the excitation of hybrid plasmons provided that the Fermi energy of electronic system is comparable to the SO phonon energy \cite{hwang2014}. The angle-resolved reflection electron energy-loss spectroscopy is a suitable tool for measuring these coupled modes.\cite{hwang2010} The electron-SO phonon coupling which can be modeled by the Fr$\ddot{\mathrm{o}}$hlich interaction has been studied for several 2D materials starting from the GaAs-based conventional 2D electron gas\cite{jalabert}, to the more recent ones such as graphene\cite{hwang2013surface, hwang2010, hwang2014, SOgraphene} and phosphorene\cite{phosphorene} theoretically and experimentally\cite{Fei,Liuexperiment,Koch,coupledgrapheneexperiment}.

A many-body quantity which is of special experimental interest in electron gas systems is the quasiparticle lifetime or inelastic scattering rate. For a doped gapped graphene, the quasiparticle lifetime calculations have been performed and a reduction in its value by increasing the gap has been obtained. \cite{lifetimegappedgraphene2}  Also, the single-particle relaxation time of silicene in the presence of neutral and charged impurities has been investigated and compared with its transport relaxation time. \cite{lifetimesilicene} The inelastic scattering rate is related to the single-particle level broadening and can be calculated from the imaginary part of the self-energy that contains information about different interactions. As a consequence of electron-SO phonon coupling, it is expected that the inelastic quasiparticle scattering rate will change, qualitatively or/and quantitatively. This can be well understood by noting that the phonon-electron interaction may open new channels for the electron scattering via the hybrid plasmon modes damping or may suppress the existing channels. The change of scattering (damping) rate due to the electron-SO phonon coupling in a GaAs-based quasi 2D electron gas \cite{jalabert} and monolayer and bilayer graphene \cite{hwang2014} has been studied at both low and high electron densities. It was shown that in the strong coupling regime, an additional decay channel for quasiparticles can be opened which results in a new abrupt jump in the scattering rate.    

In this article, we study the interaction of electrons with SO phonons of HfO$_2$, as a polar substrate, and obtain its effect on the collective plasmon excitations in valley-spin-polarized (VSP) silicene under an exchange field and at zero temperature. We start with the dynamical dielectric function of the coupled system within the generalized RPA which includes both the Coulomb electron-electron and Fr$\ddot{\mathrm{o}}$hlich electron-phonon couplings. We calculate the coupled plasma oscillations in four different spin and valley states. We show that the available regions for the single-particle excitations and consequently the hybrid plasmonic modes, depend considerably upon the spin and valley degrees of freedom in each electronic state. Interestingly, we find that while both coupled modes are well-defined in one valley, it is likely that one of them will be completely damped and disappeared in the other valley. Furthermore, we compute the intra-subband inelastic scattering rate of quasiparticles from the $G^{0}W$ approximation of the electron-SO phonon coupled self-energy for each valley and spin index. 
Since the interaction between SO phonons and electrons in VSP silicene under an exchange field leads to dissimilar hybrid plasmon modes and quasiparticles scattering rate in  different VSP subbands, this suggests that the electron-SO phonon coupling in such systems has a potential to be used in valleytronics and spintronics applications.

The rest of the paper is organized as follows: In the next section, we present the theory and our numerical results together with extensive discussions. Finally, the highlights of this work are summarized in conclusion.

\section{Theory and Results}
The structure of silicene can be modeled as a 2D honeycomb lattice which is slightly buckled and can be treated as a combination of two sublattices, displaced from each other by 0.46 $A^{\circ}$ \cite{bandgap1,bandgap2}. The band structure of silicene has an intrinsic band gap, $2\Delta_{soc}=7.8$ meV \cite{spin1}, as a result of non-zero SOC which arises from the buckling. 

\begin{figure}
	\centering
	\includegraphics [width=1.1\linewidth]{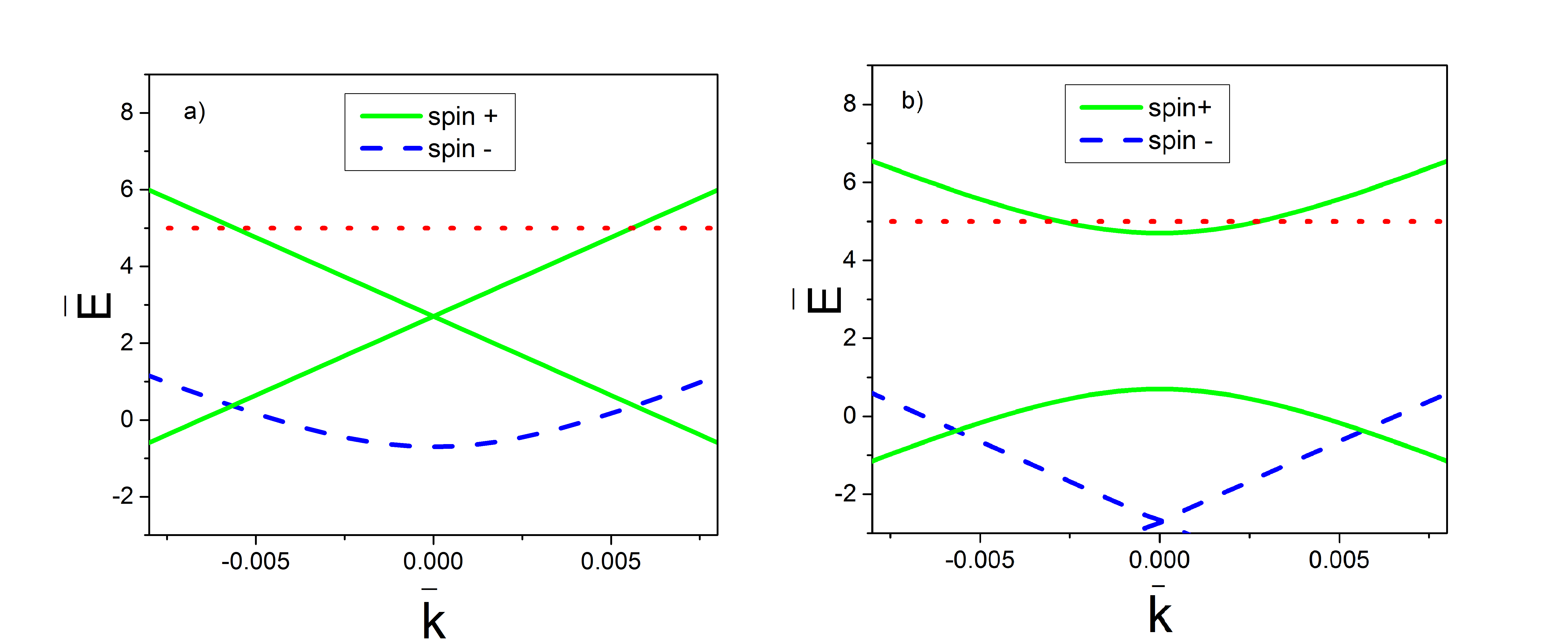}  
	
	\caption{(Color online) Band structure of VSP silicene around the Dirac points with $M=2.7\Delta_{soc}$ for two different valleys: a) K$_1$ and b) K$_2$. The solid green line corresponds to spin-up and dashed blue line shows spin-down states. The red dotted line shows the scaled chemical potential, $\mu_{0}/\Delta_{soc}$. Here, we define the dimensionless energy and wave vector, $\bar{E}=E/\Delta_{soc}$ and $\bar{k}=ka$ with $a=3.89$\,\AA.}
	\label{fig:bandstrucure}
\end{figure}
The effective low energy tight-binding Hamiltonian of silicene in the presence of external static electric and magnetic fields near the Dirac points, K$_1$ and K$_2$, is given by \cite{peters,VHE,soc}
\begin{equation}
\begin{aligned}
H_{\eta s}=
\begin{bmatrix}
\Delta_{\eta s}+s M & \hbar v_{F} (\eta k_{x}-{ik_{y}})   \\
\hbar v_{F} (\eta k_{x}+{ik_{y}}) & -\Delta_{\eta s}+s M 
\end{bmatrix} \ ,
\end{aligned}
\label{eq:hamiltoni}
\end{equation}
where $s={\pm}1$ and ${\eta}={\pm}1 $ are the spin and valley indexes and $k_{x}$ and $k_{y}$ being the Cartesian components of 2D wave vector $\mathbf{k}$. $\Delta_{\eta s}=|{\eta} s \Delta _{soc}-\Delta_{z}|$ is half of the spin and valley dependent band gap which is controlled by $\Delta_{z}$, the on-site potential difference between two sublattices due to the external electric field. Also, ${\Delta_{soc}}$ is the intrinsic SOC energy and ${v_F}=5{\times}10^{5}$ m/s denotes the Fermi velocity. 
In this Hamiltonian, the effect of Rashba-type SOC is ignored thus the spins are independent from each other. Similarly, since the two valleys are not coupled and the inter-valley processes are not included, the Hamiltonian can be written in such reduced form. 
According to the above Hamiltonian, the energy spectrum can be obtained as 
\begin{equation}
E^{s}_{\eta}=sM+\lambda\sqrt{{{\hbar^{2}}}{v_F^{2}}k^{2}+\Delta^{2}_{{\eta}s}},
\label{eq:E}
\end{equation}
where $\lambda={\pm}1$ is the band index. As mentioned earlier, by setting $ \Delta_{z}=\Delta _{soc}$, the valley-spin polarization is achieved in silicene. 

In Fig.\,\ref{fig:bandstrucure}, the band structures of VSP silicene at two Dirac valleys around K$_1$ and K$_2$, are shown. It is worth pointing out that while $\Delta_{z}$ changes the band gap between the same spin-polarized subbands, the exchange field, $M$, equally displaces these same spin subbands.\cite{peters} As it can be observed, when both the electric and exchange fields are applied to silicene, two spin components are displaced in opposite directions and this effect is vice versa in different valleys. The Fermi level is controlled by the exchange field, hence the density of electrons in each subband depends on the magnitude of this field. In the rest of this paper, we focus on VSP silicene placed on an insulator substrate under the conditions given in Fig.\,\ref{fig:bandstrucure} and consider the two important electron-electron and substrate electron-SO phonon interactions.

\begin{figure*}[ht!]
	\centering
	\includegraphics [width=0.8\linewidth]{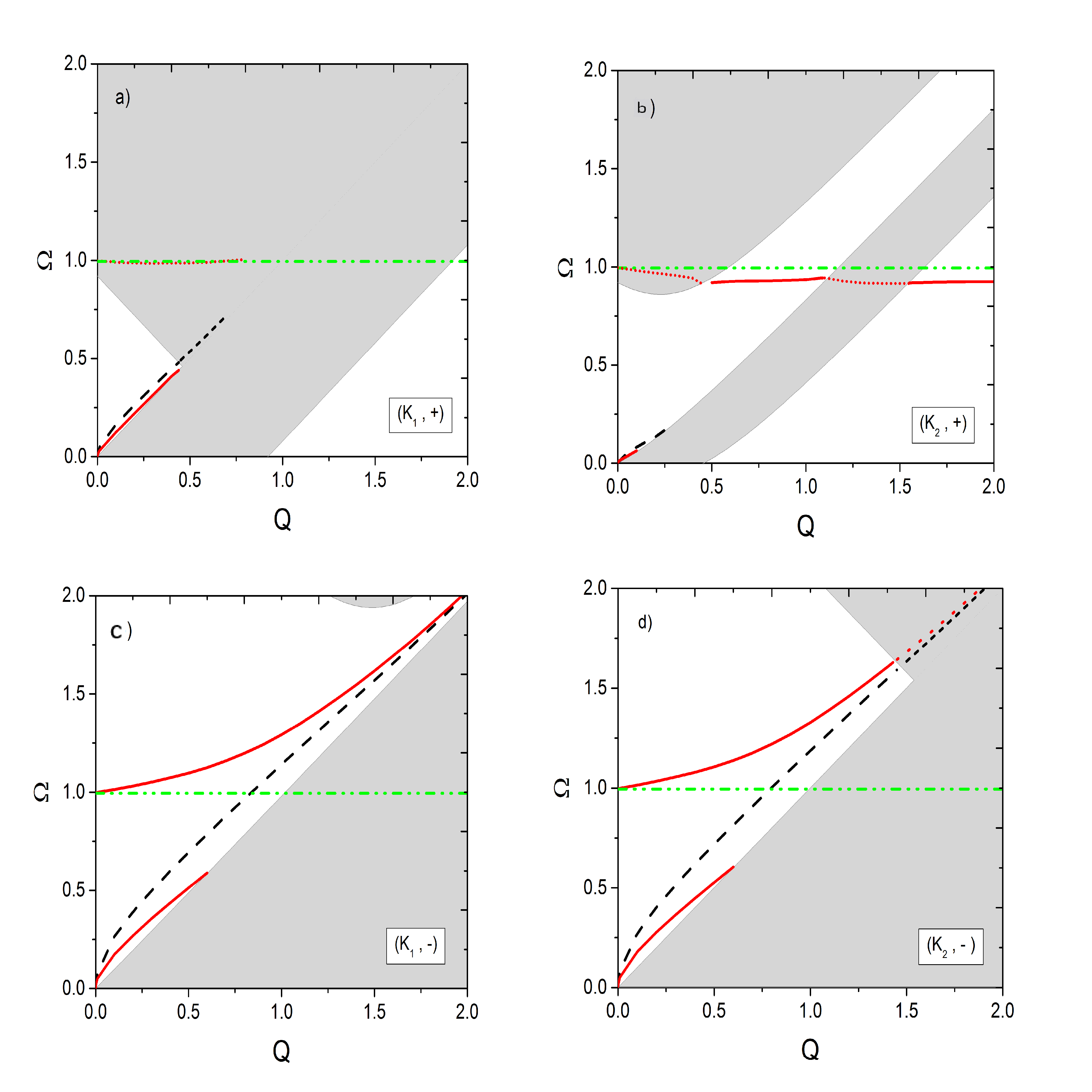}  
	\caption{(Color online) Uncoupled (black curves) and plasmon-SO phonon coupled (red curves) modes of VSP silicene on  HfO$_{2}$ as a substrate. The shaded areas correspond to the SPE continua.  Panels (a)-(d) show the calculations for different spin-valley subbands. The black short-dashed (red dotted) line of the uncoupled (coupled) plasmon branch represents the damped modes. The horizontal green dash-double-dotted line is the uncoupled SO phonon dispersion.  $Q$ and $\Omega$ are dimensionless parameters defined as: $Q=\hbar v_F q/ \mu_0$ and $\Omega=\hbar \omega/\mu_0$. }
	\label{fig:plasmon}
\end{figure*}
To include the interactions between the electrons, we use the following Hamiltonian that describes the intra-valley electron-electron interaction \cite{Mahan}
\begin{equation}
\begin{aligned}
H_{e-e}=\frac{1}{2}\sum_{\mathbf{pqk}}\sum_{\eta s_{1} s_{2}}  v_{c}(q) a^{\dagger}_{\mathbf{k+q},\eta s_{1}} a^{\dagger}_{\mathbf{p-q}, \eta s_{2}} a_{\mathbf{p},\eta s_2} a_{\mathbf{k},\eta s_{1}} \ ,
\label{eq:Hee}
\end{aligned}
\end{equation}
where $ a^{\dagger} $ ($a$) is the electron creation (annihilation) operator and $v_{c}(q)=2\pi{e^{2}}/{\kappa_{\infty} q}$ is the Coulomb potential with $\kappa_{\infty}$ being the background high-frequency dielectric constant. It is well-known that the presence of phonons in electron gas systems leads to the electron intra- or inter-valley scattering. \cite{kioseoglou2016optical,gunst2016first,prunnila2005intervalley,phonon}
\subsection{Hybrid plasmon-SO phonon modes}
\begin{figure*}
	\centering
	\includegraphics [width=0.93\linewidth]{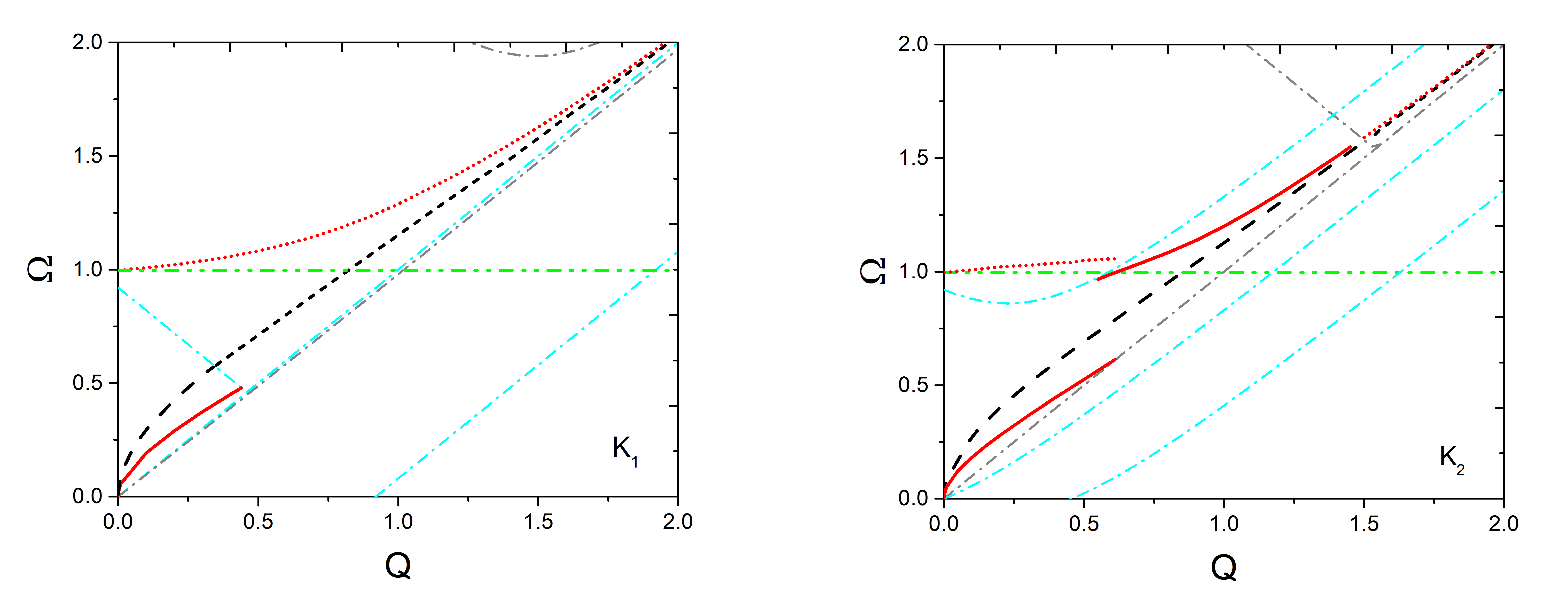}  
\caption{(Color online) Uncoupled (black curves) and plasmon-SO phonon coupled (red curves) modes of VSP silicene on  HfO$_{2}$ in two different valleys: K$_1$ valley on the left and K$_2$ valley on the right panels. The dash-dotted blue (gray) lines represent the boundaries of SPE region for spin-up (down) state in each valley as displayed in Fig.\,\ref{fig:plasmon} and the black short-dashed (red dotted) line of the uncoupled (coupled) plasmon branch represents the damped modes. The horizontal green dash-double-dotted line is the uncoupled SO phonon dispersion. Dimensionless parameters $Q=\hbar v_F q/ \mu_0$ and $\Omega=\hbar \omega/\mu_0$ are used.}
\label{fig:k1k2plasmon}
\end{figure*}
\begin{figure*}
	\centering
	\includegraphics [width=0.90\linewidth]{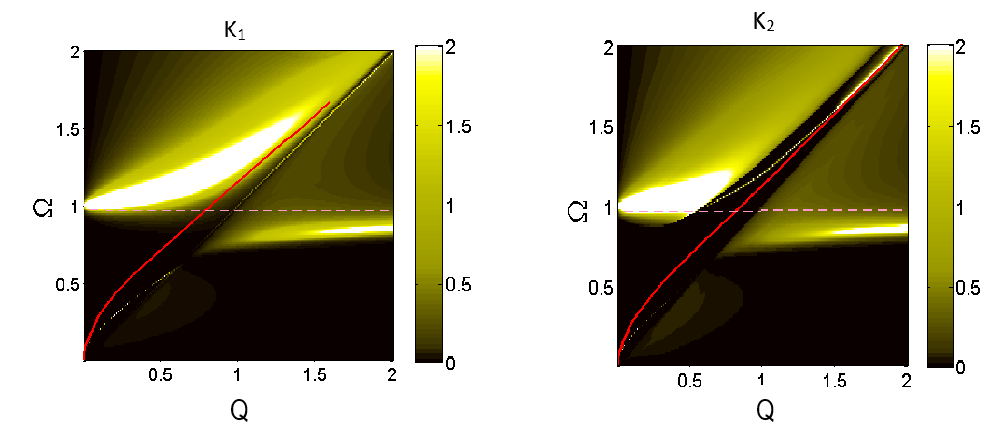}  
	\caption{(Color online) Valley-dependent energy loss function of VSP silicene on HfO$_{2}$ in K$_{1}$ valley (left panel) and K$_{2}$ valley (right panel). The solid red line indicates the uncoupled plasmons and horizontal line shows $\hbar\omega_{SO}/\mu_0$. Dimensionless parameters $Q=\hbar v_F q/ \mu_0$ and 
		$\Omega=\hbar \omega/\mu_0$ are used.}
	\label{fig:k1k2lossfunction}
\end{figure*}
The electrons in silicene on polar substrate can be affected by the long-range electric field of the SO phonons of substrate at long wavelengths. This phonon-electron coupling is a source of the intra-valley scattering. In order to investigate the effect of SO phonons, it is sufficient to calculate the total dielectric function of the system which includes the contributions from both electrons and phonons. In this paper, we assume that silicene is placed on HfO$_2$, a polar insulator, so carriers in silicene interact with substrate SO phonons through the long-range Fr$\ddot{\mathrm{o}}$hlich coupling. This interaction is given by\cite{Mahan}
\begin{equation}
\begin{aligned}
H_{e-ph}=\sum_{\mathbf{kq}}\sum_{\lambda \lambda^{'}}\sum_{\eta \sigma}M^{\lambda \lambda^{'}}_{kq} a^{\dagger}_{\mathbf{k+q},\eta \sigma} a_{\mathbf{k},\eta \sigma}(b_{\mathbf{q}}+b^{\dagger}_{\mathbf{-q}}).
\label{eq:eq4}
\end{aligned}
\end{equation}
Here, $b^{\dagger}_{\mathbf{-q}}$ and $b_{\mathbf{q}}$ are defined as the creation and annihilation operators for surface phonons, and the matrix elements of the phonon-electron interaction $ M^{\lambda \lambda\prime}_{kq}$ can be written as 
\begin{equation}
\begin{aligned}
M^{\lambda {\lambda^{'}}}_{kq}=M_{0}({q})F^{\dagger}_{{k+q},\lambda}F_{{k},\lambda^{'}} \ ,
\label{eq:M}
\end{aligned}
\end{equation} \\
where $F_{\mathbf{k}}$ is the Bloch spinor for a massive Dirac fermion \cite{spinor}
\begin{equation}
F_{k,\lambda}=  
\left(\begin{array}{c} \cos{(\theta^{\eta s} _{k}/2)} \\ \eta \sin{(\theta^{\eta s} _{k}/2)}e^{i\eta\varphi} \end{array}\right) 
\ ; \  \lambda=+1 ,
\end{equation}
\begin{equation}
F_{k,\lambda}= 
\left(\begin{array}{c} -\eta \sin{(\theta^{\eta s} _{k}/2)}e^{-i\eta \varphi} \\ \cos{(\theta^{\eta s} _{k}/2)} \end{array}\right) 
\ ; \  \lambda=-1 .
\end{equation}\\
Here, we have $\cos{\theta^{\eta s}_k}=\Delta_{\eta s} /{\sqrt{(\hbar v_F k)^2+\Delta_{\eta s}^{2 }}}$ and 
$\sin{\theta^{\eta s} _{k}}=|\hbar v_F k|/{\sqrt{(\hbar v_F k)^2+\Delta_{\eta s}^{2 }}}$. In addition, $M_0(q)$ is expressed as
\begin{equation}
\begin{aligned}
{[M_0(q)]^{2}}=v_{c}(q) e^{-2qd}\frac{\omega_{SO}}{2}\kappa_{\infty}\left[\frac{1}{{\kappa_\infty}+1}-\frac{1}{{\kappa_0}+1}\right].\\
\label{eq:M0}
\end{aligned}
\end{equation}
In the above equation, $\omega_{SO}$ is the SO phonon frequency, $d$ denotes the distance between silicene and polar substrate and $\kappa_{0}$ is the zero-frequency dielectric constant.
The interaction between the charge-density collective modes of electrons and SO phonons is a macroscopic coupling and can be investigated through the total dynamical dielectric function, ${\varepsilon_{t}(\mathbf{q},\omega)}$.
Knowledge of the total dielectric function allows us to obtain the collective and single-particle excitations in the system.
Within the generalized RPA framework, the total dynamical dielectric function which includes both the electron-electron and electron-SO phonon interactions is expressed as\cite{Mahan}

\begin{equation}
\\ \varepsilon_{t}(\mathbf{q},\omega)=1-\frac{2{\pi}e^{2}}{\kappa_{\infty}q}\Pi_{0}(\mathbf{q} ,\omega) -\frac{M^{2}_{0}(q)D_{0}(\omega)}{v_{c}(q)+M^{2}_{0}(q)D_{0}(\omega)}.
\label{eq:dielec}
\end{equation}
Here, $D_{0}(\omega)$ is the bare SO phonon propagator which is defined as
\begin{equation}
D_{0}(\omega)=\frac{2\omega_{SO}}{\omega^{2}-\omega^{2}_{SO}},
\label{eq:9}
\end{equation}
and $\Pi_{0}(\mathbf{q},\omega) $ is the non-interacting density-density response (polarization) function obtained from the bare bubble diagram.
According to Eq. ({\ref{eq:hamiltoni}}), two spins and two valleys are independent of each other in silicene, so the total polarization function $\Pi_{0}(\mathbf{q},\omega)$ is the sum of four independent terms\cite{peters}:
\begin{equation}
\Pi_{0}(\mathbf{q},\omega)=\sum_{\eta=\pm{1}}\sum_{{s}=\pm{1}}\Pi^{\eta{s}}_{0}(\mathbf{q},\omega),
\label{eq:pol0}
\end{equation}
where the spin- and valley-dependent density-density response function is expressed as
\begin{equation}
\begin{aligned}
\Pi^{\eta{s}}_{0}(\mathbf{q},\omega)=&\int\frac{d^{2}\mathbf{k}}{(2\pi)^{2}}\sum_{\lambda \lambda^{\prime}=\pm{1}}f^{\lambda \lambda^{\prime}}_{\eta{s}} (\mathbf{k,k+q}) \\
&\times \frac{n_{F}(\lambda E^{\eta{s}}_{\mathbf{k}})-n_{F}(\lambda^{\prime} E^{\eta{s}}_{\mathbf{k+q}})}{\hbar \omega+\lambda E^{\eta{s}}_{\mathbf{k}}-\lambda^{\prime}  E^{\eta{s}}_{\mathbf{k+q}}+i\delta}.
\end{aligned}
\label{eq:pol}
\end{equation}

\begin{figure*}
	\centering
	\includegraphics [width=0.92\linewidth]{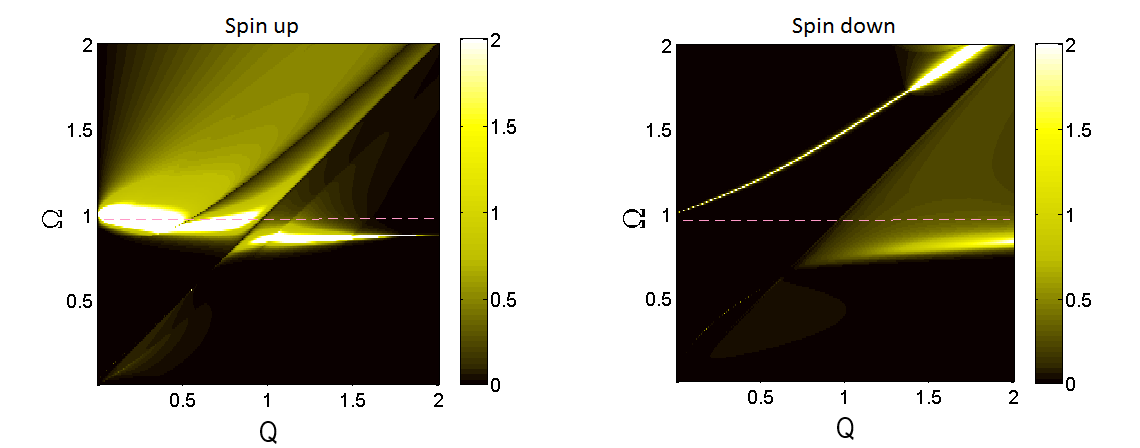}    
	\caption{(Color online) Spin-dependent energy loss function of VSP silicene on HfO$_{2}$ for spin-up (left panel) and spin-down (right panel) subbands. The horizontal line shows $\hbar\omega_{SO}/\mu_0$. Dimensionless parameters $Q=\hbar v_F q/ \mu_0$ and $\Omega=\hbar \omega/\mu_0$ are used.}	
	\label{fig:spinlossfunction}
\end{figure*}
Here, $E^{\eta{s}}_{\mathbf{q}}=[\hbar^{2}v_{F}^{2}q^{2}+\Delta^{2}_{\eta{s}}]^{1/2}$ and $n_{F}(E_\mathbf k)=1/(1+\exp[(E_\mathbf k-\mu_{s})/k_{B}T])$ is the Fermi-Dirac distribution with $\mu_{s}=\mu_{0}-s{M}$, $k_{B}$ and $T$ being the exchange field dependent Fermi level, Boltzmann constant and temperature. The form factor of silicene, ${f^{\lambda \lambda^{'}}_{\eta{s}}}$, is calculated from the overlap between the eigenstates \cite{Pyatkovskiy}
\begin{equation}
\begin{aligned}
f^{\lambda \lambda^{\prime}}_{\eta{s}}=&{|F^{\dagger}_{\lambda \mathbf{k} } F_{\lambda^{\prime}\mathbf{k+q}}|}^{2}\\
=&\frac{1}{2}[1+\lambda \lambda^{\prime} \frac{{\hbar}^2{v_F}^{2}\mathbf{k}(\mathbf{k+q})+\Delta^{2}_{\eta{s}}}{E^{\eta{s}}_{\mathbf{k}}E^{\eta{s}}_{\mathbf{k+q}}}] .
\end{aligned}
\label{eq:struc}
\end{equation}
The polarization function of silicene has been analytically calculated at zero temperature \cite{peters,tabert}.
To investigate the coupling between substrate phonons and plasmons of silicene, we calculate hybrid collective excitation modes which are obtained from the zeros of total complex dielectric function (Eq.\,(\ref{eq:dielec})). In 
Fig.\,\ref{fig:plasmon}, we show the uncoupled and coupled plasmon-phonon modes of silicene for four different conduction band states of VSPM phase, {\it{i.e.}} (K${_1} ,\pm)$, (K${_2} ,\pm)$, in the presence of electric and exchange fields with HfO$_{2}$ as the polar substrate. It should be noted that while the total dielectric function is introduced as $1-v_c(q)\Sigma_{\eta s}\Pi_{\eta s}$, we calculate the plasmon modes in Fig.\,\ref{fig:plasmon} from $1-v_c(q)\Pi_{\eta s}$ to study the plasma oscillations behavior and contribution from each subband, separately. The following parameters have been used in the calculations presented in this paper: $ \omega_{SO}=19.42$ meV, ${\kappa_0=22}$, ${\kappa_{\infty}=5.03}$\cite{parameter}, $d=5A^{\circ}$,  $\Delta_{z}=\Delta_{soc}$, $M=2.7 \Delta_{soc}$\cite{peters}, $\mu_{0}=5 \Delta_{soc}$ and $T=0$. Also, we define the dimensionless parameters $\Omega=\hbar \omega/\mu_{0}$ and $Q=\hbar v_F q/\mu_{0}$. In Fig. \ref{fig:plasmon}, the shaded areas represent the SPE region where the imaginary part of the dielectric function is non-zero.

\begin{figure*}
	\centering
	\includegraphics [width=0.93\linewidth]{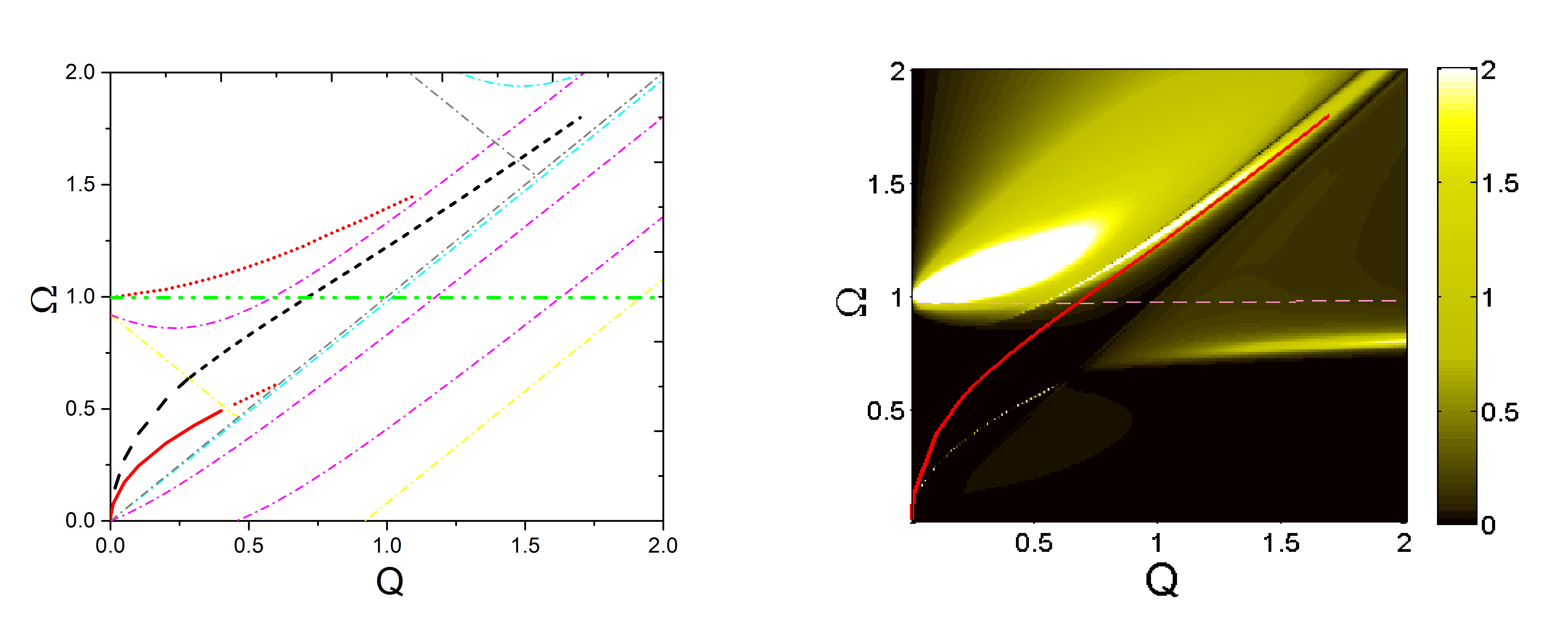} 
	\caption{(Color online) Uncoupled (black curves) and plasmon-SO phonon coupled (red curves) modes in VSP silicene on HfO$_{2}$ with dash-dotted yellow, blue, violet and gray curves being the boundaries of the SPE regions associated to (K${_1},+$), (K${_1},-$), (K${_2},+$) and (K${_2},-$) subbands, respectively as displayed in Fig.\,\ref{fig:plasmon} and with the black short-dashed (red dotted) line of the uncoupled (coupled) plasmon branch showing the damped modes (left panel) and the corresponding electron loss function (right panel). The horizontal line is the uncoupled SO phonon dispersion. Dimensionless parameters $Q=\hbar v_F q/ \mu_0$ and $\Omega=\hbar \omega/\mu_0$ are used.}
	\label{fig:totalplasmon} 
\end{figure*}

In the phonon-plasmon coupled system, there are two plasmon branches called phonon-like (high energy branch) and plasmon-like (low energy branch) .As illustrated in this figure, the plasmon modes of the uncoupled state and the plasmon-like branch of the coupled state get damped at a larger momentum in (K$_1,+$) valley with respect to the (K$_2,+$) case. In the former case, the plasmon energies have greater values corresponding to a higher electron density in this state. Moreover, the plasmon-SO phonon coupling is clearly weak in these two cases, as the phonon-like (plasmon-like) branch lies very close to the uncoupled phonon (plasmon) modes. It is worth pointing out that the phonon-like branch of (K$_1,+$) sits in the SPE continuum for any wave vector and gets easily damped, whereas in (K$_2,+$) case, some part of this branch is out of the SPE region. In contrast, the plasmon-phonon coupling for the electrons with spin down in both valleys, (K$_1,-$) and (K$_2,-$), is stronger and plasmon-like modes enter the SPE area at higher values of the wave vector because of larger electron densities in these subbands.

In order to make the difference between the two valleys clearer, we depict the valley-dependent hybrid modes in 
Fig.\,\ref{fig:k1k2plasmon}, with both spin-up and spin-down electrons. 
It can be seen that for K$_1$ valley, the inter-band SPE region is wider compared to K$_2$. As a result, there is less phase space available for the collective excitations. Moreover, the plasmon-like modes in silicene at K$_2$ coincides with SPE region at large frequency and the phonon-like branch does not lie entirely in the SPE, unlike K$_1$. 
In Fig.\,\ref{fig:k1k2lossfunction}, we plot the loss function, $-{\rm Im}[1/\varepsilon_{t}(\mathbf{q},\omega)]$ of silicene on HfO$_2$ in the presence of electron-SO phonon coupling for two distinct valleys. While the reduction in energy can be estimated by the loss function, its poles represent the dissipation via plasmonic excitations. In addition, it is well-known that decrease of electron energy in the SPE region is mainly due to electron-hole excitations and out of this region is mostly a result of the plasmons emissions. 

The existence of phonon-like modes in the intermediate energy interval can be observed from Fig.\,\ref{fig:k1k2lossfunction}. The contributions from the separate spin-up and spin-down electrons to the loss function in a coupled phonon-electron silicene system is displayed in Fig.\,\ref{fig:spinlossfunction}. There is a significant difference between the two phonon-like branches of these states. Diverse SPE channels are obtained due to their different density of states which change the plasmon peaks in the loss spectrum.

The total response of the coupled system calculated from Eq.\,(\ref{eq:dielec}) and the density plot of corresponding loss function in the presence of applied electric and magnetic fields, is plotted in Fig.\,\ref{fig:totalplasmon}. It can be seen that the phonon-like plasmon branch falls into the SPE region, so in VSPM phase of silicene only the plasmon-like modes are well-defined and undamped at long wavelengths. This is a special feature and may probably be used in valleytronic and spintronic applications.

\subsection{Quasiparticle scattering rate}
One of the single-particle quantities that is influenced by the interactions is the inelastic carriers lifetime or equivalently the scattering rate. In this part, we calculate the valley- and spin-dependent inelastic quasiparticle lifetimes of VSP silicene due to the electron-electron interaction and in the presence of long-range polar Fr$\ddot{\mathrm{o}}$hlich phonon-electron coupling within the $G^{0}W$ approximation.
The quasiparticle scattering rate is obtained from the imaginary part of the retarded self-energy by making use of the so-called on-shell approximation: \cite{guliani} 

\begin{equation}
\begin{aligned}
\frac{1}{\tau_{\lambda \eta s}(\mathbf{k})}=\left|\frac{2}{\hbar}{\rm Im} [\mathit{\Sigma}^{ret}_{\lambda \eta s}(\mathbf{k},\xi^{\eta s}_{\lambda\mathbf{k}}/\hbar)]\right| \ ,
\end{aligned}
\label{eq:time}
\end{equation}
where $\xi^{\eta s}_{\lambda\mathbf{k}}=\lambda E^{\eta s}_{\mathbf{k}}-\mu_{s}$ is the energy of a quasiparticle with respect to the spin dependent Fermi energy.

Using the $G^{0}W$ formalism for 2D electron gas systems, the self-energy of silicene for each band, valley and spin state is given by: \cite{lifetimegraphene}
\begin{equation}
\begin{aligned}
\mathit{\Sigma_{\lambda \eta s}}(\mathbf{k},i\omega_{n}) =-k_{B}T\sum_{\lambda^{\prime}}\int\frac{d^{2}\mathbf{q}}{(2\pi)^2}\mathit{f}^{\lambda\lambda^{\prime}}_{\eta s}(\mathbf{k,k+q})\\
\times \sum^{\infty}_{m=-{\infty}} \frac{v_{c}(q)}{\varepsilon_{t}(\mathbf{q},i\omega_{m})}  G^{0}_{\lambda^{\prime} \eta s}(\mathbf{k+q},i\omega_{n}+ i \omega_{m})\ . 
\end{aligned}
\label{eq:selfenergy}
\end{equation}
Here, $G^{0}_{\lambda \eta s}(\mathbf{k},i\omega_m)=1/(i\omega_m -\xi^{\eta{s}}_{\lambda \mathbf{k}}/\hbar)$ is the non-interacting Green's function with $\omega_{n}$ and $\omega_{m}$ indicating the fermionic and bosonic Matsubara frequencies, respectively.
After performing the summation over $m$ and employing the analytic continuation, i$\omega_{n}$ to $\omega+i\delta$, the obtained expression for the self-energy may be divided into two exchange and correlation terms, $\varSigma^{ret}_{\lambda \eta s}(\mathbf{k},\omega)=\varSigma^{ex}_{\lambda \eta s}(\mathbf{k})+\varSigma^{cor}_{\lambda \eta s}(\mathbf{k},\omega)$. The correlation term in $G^{0}W$ approximation itself consists of two parts, line and pole components, $\varSigma^{cor}=\varSigma^{line}+\varSigma^{pole}$. While the exchange and line parts of the self-energy are completely real, the imaginary contribution originates from the pole term of the correlation and is given by \cite{lifetimegraphene}
\begin{figure*}
	\centering
	\includegraphics[ width=0.93\linewidth] {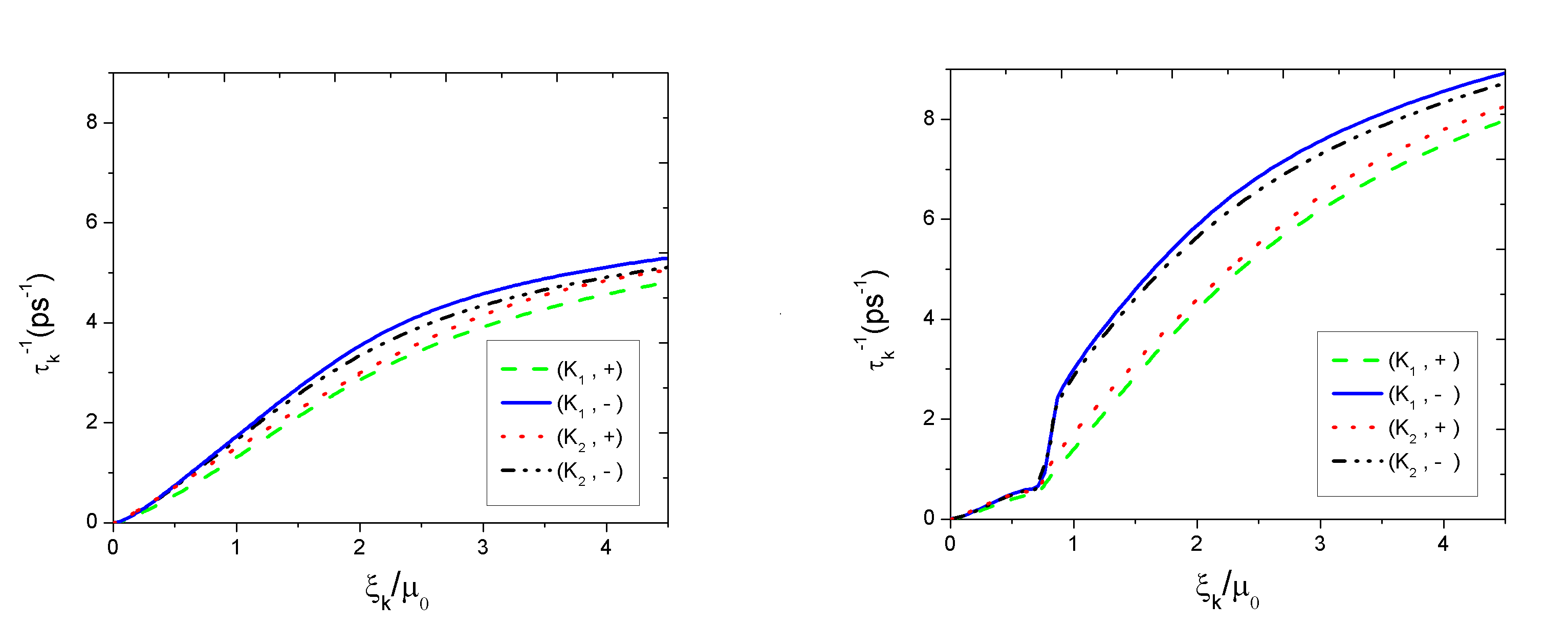}    
	\caption{(Color online) Quasiparticle scattering rate as a function of $\xi^{}_{k} /\mu_0 $, where $\xi^{}_{k}=\xi^{\eta s}_{\lambda\mathbf{k}}$ with $\lambda=1$ is the valley- and spin-dependent on-shell energy in four different subbands of VSP silicene: (K$_{1},\pm$) and (K$_{2},\pm$) for a) uncoupled and b) electron-SO phonon coupled systems.} 
	\label{fig:lifetime}
\end{figure*}
\begin{equation}
\begin{aligned}
{\rm Im}\mathit{\Sigma}^{ret}_{\lambda \eta s}(\mathbf{k},\omega)
=\sum_{\lambda^{\prime}}\int\frac{d^{2}\mathbf{q}}{(2\pi)^2}{\rm Im}[\frac{v_{c}(q)}{\varepsilon_{t}(\mathbf{q},\omega-{\xi}^{\eta s}_{{\lambda^{\prime}}\mathbf{k+q}}/{\hbar})}] \\
\times\,\mathit{f}^{\lambda \lambda^{\prime}}_{{\eta}s}(\mathbf{k,k+q})
[\ominus(\omega-\xi^{\eta s}_{{\lambda^{\prime}} \mathbf{k+q}}/\hbar)-\ominus(-\xi^{\eta s}_{{\lambda^{\prime}}\mathbf{k+q}}/\hbar)]\ .
\end{aligned}
\label{eq:imselfenergy}
\end{equation}
where $\ominus(x)$ is the Heaviside unit step function.  
A structure with separated spin states in each valley may have some available mechanisms for scattering, corresponding to the intra-subband and inter-subband excitations for each spin, among which the scattering with spin flip is negligible in the low-frequency excitations. 

In Fig.\,\ref{fig:lifetime} the valley- and spin-dependent intra-subband quasiparticle scattering rates within the $G^{0}W$ are plotted for four possible states, (K$_{1},\pm $) and (K$_{2},\pm $), of the conduction band in the absence and presence of the electron-SO phonon coupling. Here the scattering rate is dominated by two mechanisms: emission of the plasmons \cite{jalabert} and intra-subband particle-hole excitations. The sharp increase in the scattering rate curve determines a threshold energy (or equivalently a threshold wave vector) from which the plasmons enter the intra-subband Landau damping region. It can be observed from Fig.\,\ref{fig:lifetime}(a) that the scattering rates of all uncoupled states have the same behavior without any jump that indicates the contribution from the plasmon damping is negligible. Also, the calculated results for distinct subbands are slightly different which can be related to the different band structure and electronic population of these states.   
For the gapless graphene system the scattering rate is only due to the intra-band SPE process\cite{lifetimegraphene}, and in the cases of gapped and bilayer graphene the plasmon emission is the dominant decay process \cite{lifetimegappedgraphene2,hwang2014}. In the VSP silicene, however, there is a coexistence of both the gapless and electric-induced gapped states in each valley so the situation is rather more complicated. On the other hand, in VSP silicene under an exchange field, different subbands contain unequal electron densities (see 
Fig.\,\ref{fig:bandstrucure}) that may affect the magnitude of the scattering rate and the threshold energy for the plasmon-like modes emission. \cite{hwang2014} As it is shown in Fig.\,\ref{fig:bandstrucure}, the conduction band electron concentration of the spin-up state is lower than that of the spin-down state in both valleys which leads to smaller values for the corresponding quasiparticle lifetimes. In addition, due to the quadratic form of (K$_{1(2)},-(+)$) subband and consequently different plasmon dispersions and intra-subband SPE regions, its scattering rate curves lies higher than that of the (K$_{2(1)},-(+)$) state which has a linear electronic band structure. 
In the coupled system, the change of plasmon dispersions is responsible for the new behavior of the quasiparticle lifetime. 

The effect of plasmon-SO phonon coupling on the quasiparticle scattering rate is illustrated in Fig.\,\ref{fig:lifetime}(b). In this case, the values of scattering rate increase as the plasmons enter the intra-subband SPE, especially at high energies. A sharp increase in the scattering rate for (K$_{1},-$) and (K$_{2},-$) valleys mainly arises from the damping of plasmon-like modes which begins at an intermediate wave vector. This new channel for scattering which was almost absent in the uncoupled system results in a clear separation between the spin-up and spin-down states.

\section{conclusion}
In this work, we have studied the plasmon-SO phonon coupling in monolayer silicene on HfO$_2$ as a polar substrate in the presence of perpendicularly applied electric and exchange fields and obtained the coupled plasmon oscillations. We have considered the important VSPM phase of silicene to be able to study the valley- and spin-polarized behavior of hybrid plasmons in different subbands. In order to determine the coupled plasmon-like and phonon-like branches, we have calculated the dynamical dielectric function of the system within the generalized RPA, considering both the Coulomb electron-electron and Fr$\ddot{\mathrm{o}}$hlich electron-phonon interactions. Our results have shown that the hybrid plasmon dispersions behave rather differently in separate spin-valley-locked states, for instance, the completely damped high energy (phonon-like) branch in one state is well-defined in the other subbands. 

We have also illustrated the coupled plasmon modes for subbands with the same valley and same spin indeces. It has been found that the spin-down states with higher electron densities, exhibit a stronger electron-SO phonon coupling with respect to the spin-up states.
 
In addition, we have computed the inelastic scattering rate of VSP silicene in the presence of coupling between the electrons and SO phonons of the substrate and compared our results with those obtained for the uncoupled system. We have used the $G^{0}W$ approximation to calculate the imaginary part of the quasiparticle self-energy in each valley and spin states. It turns out that while the behavior of scattering rate is almost the same for all valley and spin states in the case of uncoupled system, the situation for coupled system is different and an explicit dependence on the spin index is observed. The scattering rates of the spin-down subbands display a sharp increase as a consequence of the plasmon-like mode's damping due to entering the intra-subband SPE region at an intermediate wave vector.      
The measurable effects obtained in this work suggest potential applications in valleytronics and spintronics.

\acknowledgments{B.T. thanks TUBA for support.}

\end{document}